# FPGA Implementation of a Fixed Latency Scheme in a Signal Packet Router for the Upgrade of ATLAS Forward Muon Trigger Electronics

Jinhong Wang, Xueye Hu, Thomas Schwarz, Junjie Zhu, J.W. Chapman, Tiesheng Dai, and Bing Zhou

***Abstract*—We propose a new fixed latency scheme for Xilinx gigabit transceivers that will be used in the upgrade of the ATLAS forward muon spectrometer at the Large Hadron Collider. The fixed latency scheme is implemented in a 4.8 Gbps link between a frontend data serializer ASIC and a packet router. To achieve fixed latency, we use IO delay and dedicated carry in resources in a Xilinx FPGA, while minimally relying on the embedded features of the FPGA transceivers. The scheme is protocol independent and can be adapted to FPGA from other vendors with similar resources. This paper presents a detailed implementation of the fixed latency scheme, as well as simulations of the real environment in the ATLAS forward muon region.**

***Index Terms*— FPGA, serial link, fixed latency, changeable delay tuning**

## I. Introduction

Following the first phase of data taking of the Large Hadron Collider (LHC) in 2011-2012 and the discovery of the Higgs boson, major upgrades of the ATLAS detector are now underway. Among other improvements, the muon endcap spectrometer will be replaced by the so-called "New Small Wheels" (NSW), a combination of Micromegas chambers and small-strip thin-gap chambers (sTGC) [1]. This work pertains to the trigger electronics used for the sTGC detector, and more specifically to achieving a deterministic latency serial link between the ASIC-based frontend part and the FPGA-based signal packet router. A block diagram of the sTGC trigger electronics is shown in Fig. 1. Signals from the sTGC strip detectors will be received by a VMM, an Amplifier-Shaper-Discriminator (ASD) ASIC [2], and then processed by the Trigger-Data-Serializer (TDS) ASIC [3] before being transmitted via twinax cables to the signal packet Router [4] on the rim of the NSW detector. The Router is an FPGA-based processing circuit. It handles all incoming traffic from TDS for strips and serves as a switching yard between incoming active TDS signals and the optical outputs sent to the segment-finding circuits in the ATLAS underground counting room (USA15). Transmission between TDS and Router is done via serial copper links up to 4 meters in length at 4.8 Gbps. The latency of the serial links must be low and deterministic in the case of power cycles or resets. Deterministic latency is preferable in serial transmission applications by the NSW trigger system, where the receiving end requires a predictable arrival time of information. The TDS is a custom ASIC with built-in fixed latency in serialization [5]. This work is concentrated on implementation of a low and fixed latency serial link in the Router, which utilizes a Xilinx Artix-7 FPGA to de-serialize the TDS serial stream with GTP transceivers.

There are several previous works on fixed latency links with the embedded transceivers inside an FPGA [6-8]. In [6-7], the authors pointed out that the latency variation of GTP transceivers at the receiver end originates from the uncertainty of user clock phases from the clock and data recovery circuit (CDR). They explored the architecture of the Xilinx GTP transceivers for fixed latency, and achieved a fixed latency link using the embedded RX byte and word alignment feature of Xilinx GTP transceivers. However, not all transceivers in FPGAs provide the phase alignment feature to users, such as [9]. In [8], the authors continued the work in [7] with a clock phase compensation feature of an external Xilinx Digital Clock Manager (DCM) to further reduce the dependency on specific features of the transceivers on the latency. However, the use of DCM for a fixed latency scheme results in significant occupation of clocking resources in FPGA as the number of links increases, and thus limits the implementation of other logic. For application in the NSW electronics, one Router needs to deal with 9 to 11 independent TDS serial outputs, which makes the DCM scheme impossible in low cost FPGAs.

In this paper, we present and implement a new scheme in an FPGA to achieve fixed latency serial links between the TDS and Router in the ATLAS NSW sTGC trigger. The scheme makes use of the general IO delays inside the FPGA and releases the dependency on internal features of transceivers.

Manuscript received on May 24th. This work is supported by the Department of Energy under contracts DESC0007857 and DE-AC02-98CH10886.

The authors are with Department of Physics, University of Michigan, Ann Arbor, MI, 48109, US. (e-mail: jinhong@umich.edu; xueyehu@umich.edu).

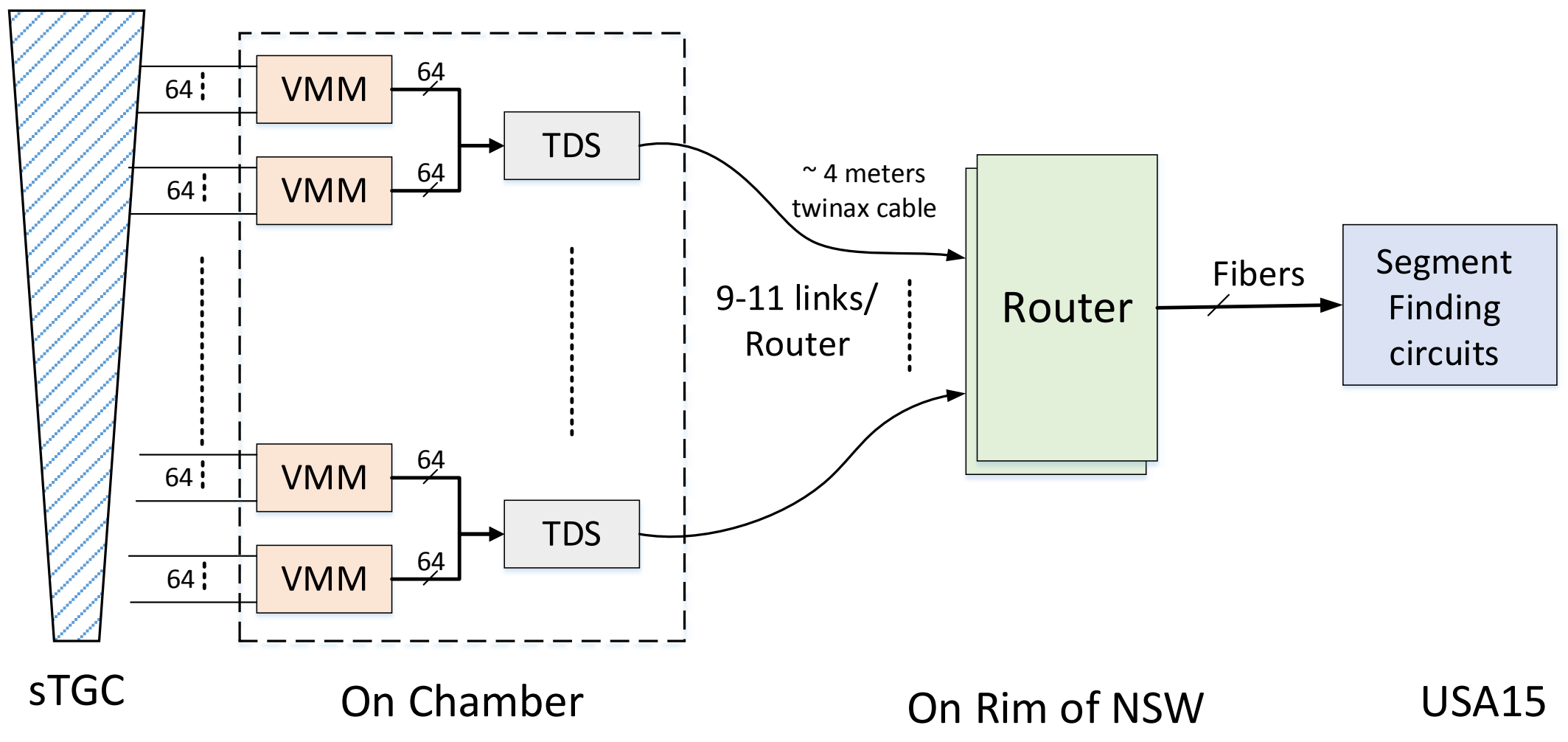

Fig. 1. Simplified block diagram of the sTGC trigger logic in NSW.

We organize the paper as follows. In Section II, we introduce the techniques used for the fixed latency implementation step by step. In Section III, we evaluate the performance with prototypes of the fixed latency scheme in the link between TDS and Router and in Section IV, we discuss the pros and cons of the scheme proposed here.

## II. Methodology

### A. *Origin of Latency Variation for a GTP RX*

Serial data stream from TDS is captured by the GTP transceiver in a Xilinx Artix-7 FPGA on the Router. A simplified block diagram of the receiving end of the Xilinx Artix-7 GTP transceiver is shown in Fig. 2. It is made up of a Physical Medium Attachment (PMA) and a Physical Coding Sub-layer (PCS). Inside the PMA, the CDR unit extracts a clock from the raw serial stream and uses it to sample the serial bit stream in a Serial-In-to-Parallel-Output (SIPO) block. SIPO buffers the sampled bits and ships the parallel bits out to the PCS via a divided version of the recovered clock (*RX UsrClk* PMA). The width of the SIPO parallel output can be set to 16, 20, 32, or 40 for GTP transceivers in the Artix-7 FPGA. For example, for a 4.8 Gbps serial link with an output width of 20 bits at SIPO, the user clock is 240 MHz. According to [7] there are 20 possible phases of the 240 MHz user clock with respect to its 4.8 Gbps serial stream. This user clock phase uncertainty leads to variations in the RX latency, and it is important to always choose an identical phase for the user clock.

There are several processing logics in the PCS, such as comma detector and 8b/10b decoder. However, these functions can be bypassed to suit users' applications. For the link between the TDS and Router, we utilize a custom protocol and bypass all processing units in the PCS to achieve a low and deterministic latency [10]. We address this protocol in the following section.

### B. *TDS and Router Serial Protocol Processing*

The proposed serial protocol between the TDS and Router is shown in Fig. 3. The width of a packet sent by a TDS ASIC is 30 bits [5]. Each signal packet starts with a 4-bit header "1010", whereas a NULL packet begins with "1100". The header is used in the Router to make a quick decision for TDS trigger data forwarding or NULL suppression. The lower 26 bits of a signal data packet are sTGC strips information, while the lower 26 bits of a NULL packet are not defined though could be utilized for other purposes. The headers are already DC balanced while the lower 26 bits are scrambled at the TDS end before being transmitted to the Router. TDS uses the scrambler scheme in IEEE Standard 802.3-2012 for 10 Gb/s Physical layer implementations (with a polynomial function of $1 + x^{39} + x^{58}$ ) [11].

Since the Artix-7 GTP RX does not support 30-bit length for the SIPO parallel output, we set the width to 20 and thus the corresponding *RX UsrClk* is 240 MHz for 4.8 Gbps. However the TDS frame clock for a 30-bit packet is 160 MHz, a circuit is required to safely transfer the serial stream from the 240 MHz *RX UsrClk* domain to the 160 MHz clock domain. To realize this, we buffer raw packets from GTP RX and pick out two TDS packets from the buffer with the packet header obtained. In total 5 consecutive RX packets (100 bits) are buffered, and the length is adequate to cover two TDS packets (60 bits) on all possible header positions. A 160 MHz clock (*Clk160*) originates from the LHC clock (40 MHz) is used as source of the TDS packet clock on the Router. *Clk160* is phase asynchronous to *RX UsrClk* and the "Clock Edge Align & Phase Compensation" unit shown in Fig. 3 aligns their rising edges. When aligned, the rising edges meet every three cycles of the *RX UsrClk*, which is the time interval of two TDS packets. Every time the rising edges of the two clocks are aligned, a *Sample* signal is generated and used to load the buffer to the "Syn. & Packet builder" block. This block performs link synchronization by recognizing the position of headers in one loaded buffer and checking the same position for headers in the following buffers. The header position is determined until a certain number of constitutive checks is passed without errors. Once the header position is confirmed, it is passed back to the "Clock Edge Align and Phase Compensation" unit. A phase-compensated 160 MHz clock (*Clk160_c*) with respect to the header position is then produced as the TDS packet clock. Meanwhile, two consecutive TDS packets are built from the buffer and a *SEL* signal is generated to multiplex the two packets to *Clk160_c* domain. There is no need to perform link synchronization again after confirmation of the header unless the link fails. The lower 26 bits of the packets are de-scrambled following the "MUX" unit.

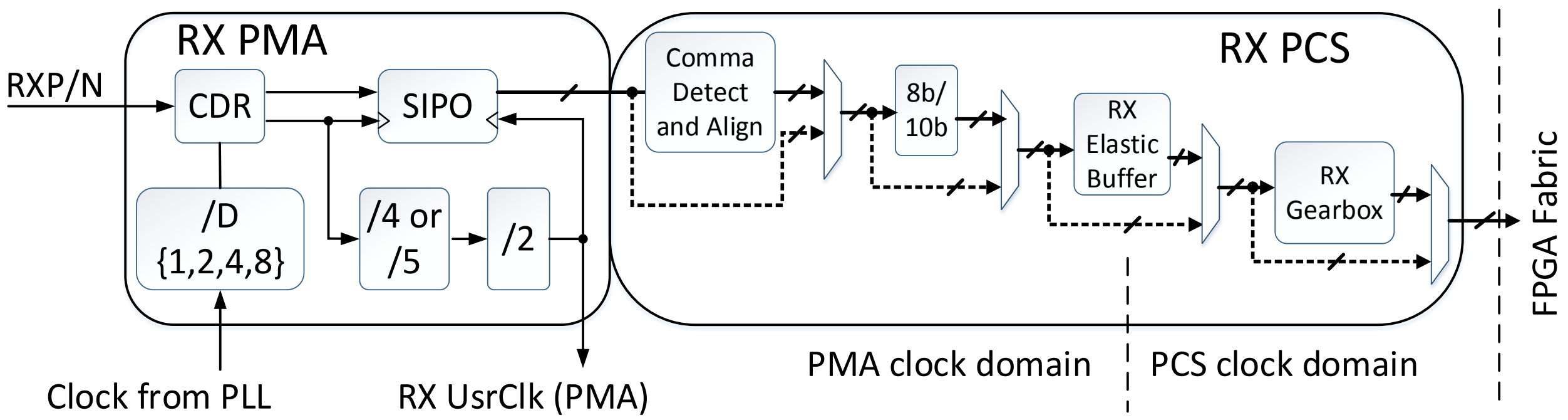

Fig. 2: Simplified diagram of the GTP RX in a Xilinx Artix-7 FPGA.

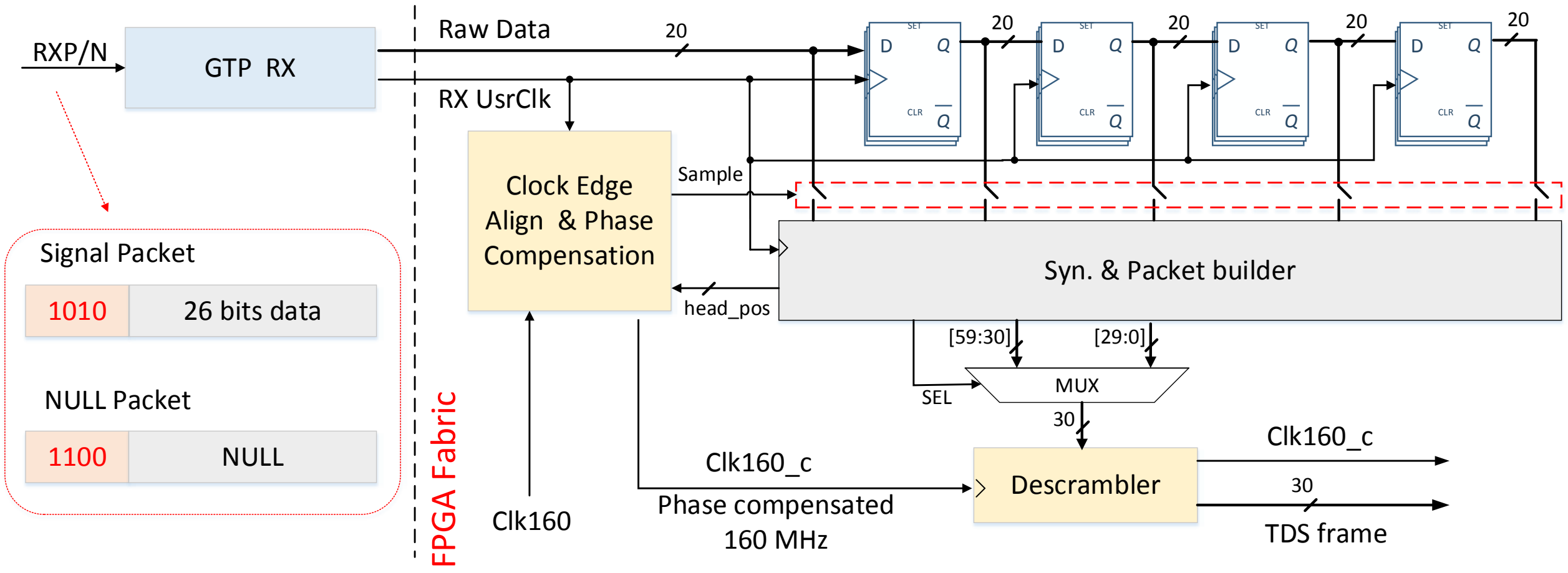

Fig. 3. TDS-Router serial protocol and corresponding processing algorithm

### C. *Clock Edge Align and Phase Compensation*

A crucial part in keeping a fixed latency in Fig. 3 is the "Clock Edge Align & Phase Compensation" unit. It aligns the rising edge of a 160 MHz clock to that of *RX UsrClk* as a reference point, from which it compensates its phase with respect to the header position to make sure an identical phase of 160 MHz is used for all header positions. We divide the procedure into two steps:

1) *Rising edge alignment of RX UsrClk and Clk160_c*
*Clk160* is phase asynchronous to *RX UsrClk* in addition to their frequency difference. *Clk160_c* is a phase-shifted version of *Clk160*. Rising edge alignment is achieved by tuning the phase of *Clk160_c* to align its rising edge to that of *RX UsrClk*.

2) *Phase compensation of Clk160_c*
This compensates for phase variation of *Clk160_c* with the packet header position from the phase alignment in Step 1.

Digital Dual Mixer Time Difference (DDMTD) [12] is widely applied for phase determination between clocks in timing systems [13]. However, DDMTD typically deals with phase difference of clocks with identical frequency, whereas for initial phase determination in Step 1, the clock frequencies are different (160 MHz versus 240 MHz). We propose to perform the alignment using the input delay resource and dedicated carry input resource in a Xilinx FPGA. The phase of *Clk160* is tuned with input delay resource and the clocking edge alignment with *RX UsrClk* is determined with the dedicated carry-in resources.

Fig. 4 shows a block diagram of the "Clock Edge Align & Phase Compensation" unit. *Clk160* is introduced to a chain of cascaded IDELAYE2s. IDELAYE2 is a 32-tap, wrap-around programmable delay primitive in every IO block of Xilinx 7 series FPGA. Though dedicated for the IO delay compensation, it can also be accessed directly for FPGA logic. The tap resolution of IDELAYE2 is 78 ps with a 200 MHz reference clock. A higher resolution of 52 ps and even 39 ps is also available with a reference clock of 300 MHz and 400 MHz respectively. For the 4.8 Gbps link between TDS and Router, the bit width is about 208.3 ps, almost covering three taps for 78 ps resolution. We consider the 78 ps mode is adequate for the link. In addition, the phase tuning circuit should cover at least the period of one TDS packet (6.25 ns) to make full phase compensation on all possible header positions. The total range that an IDELAYE2 primitive covers is 2.496 ns for 78 ps resolution. To cover a range of 6.25 ns, at least three IDELAYE2 primitives should be cascaded for 78 ps resolution, as IDELAYE2 #0-2 shown in Fig. 4.

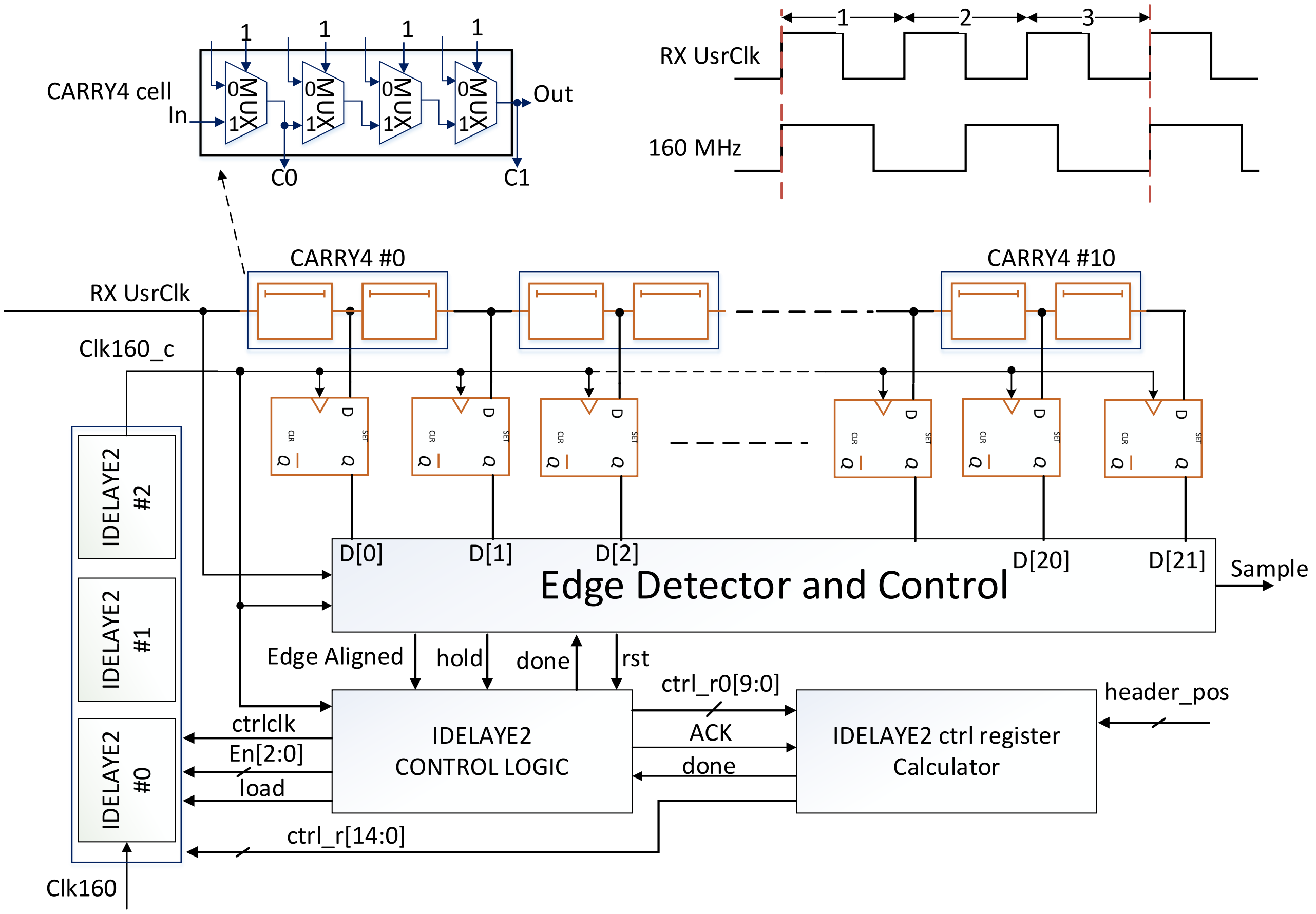

Fig. 4. TDS-Router serial protocol and corresponding processing algorithm.

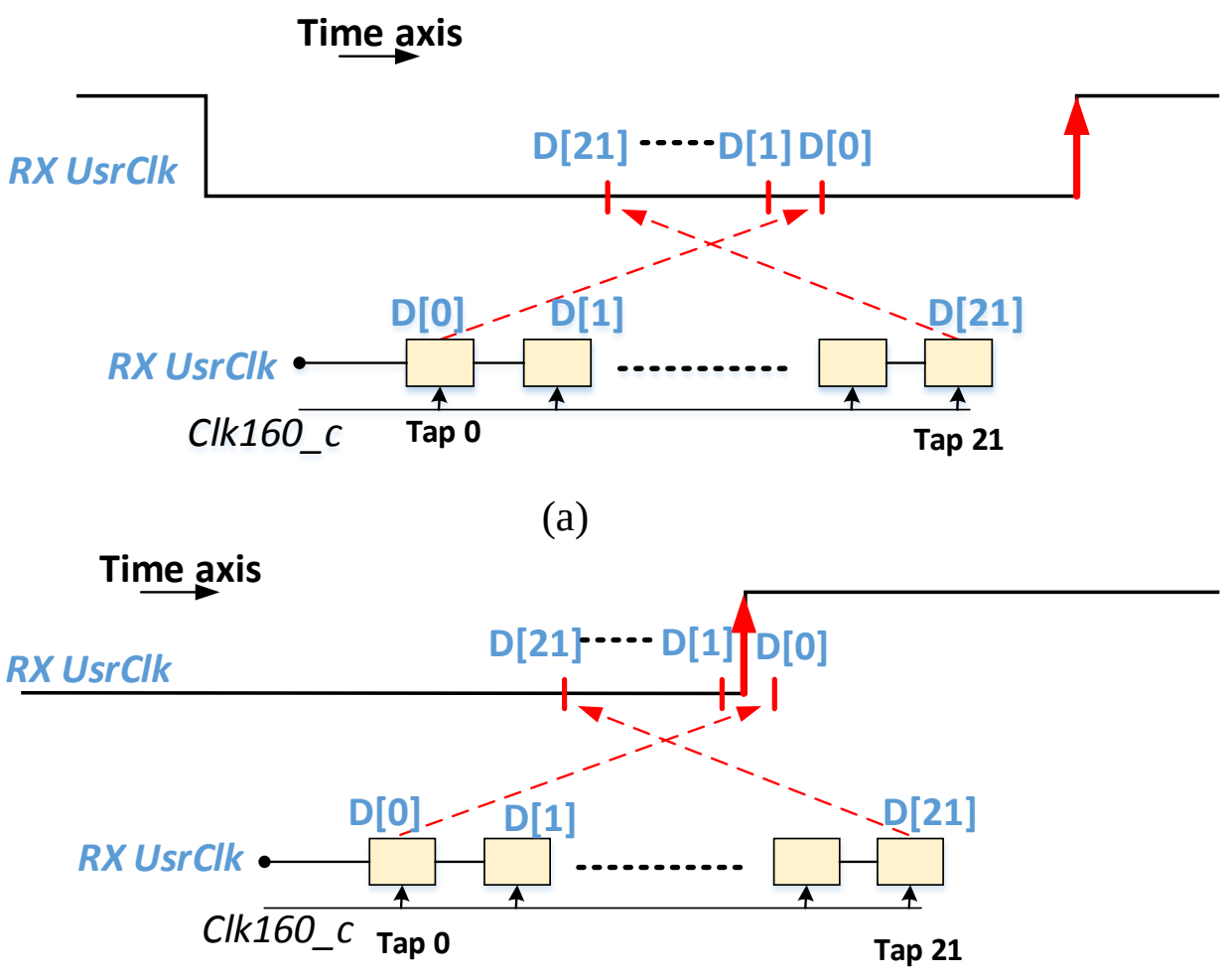

(b)

Fig. 5. Illustration of states of CARRY4 chain in rising edge alignment, where (a) shows a state before the edge of *Clk160_c* meets that of *RX UsrClk* and *D[0]-D[21]* are all zeros; and (b) shows a state the rising edges are just aligned, *D[0]-D[21]* switches to a non-zero status.

The counterpart of *Clk160*, *RX UsrClk*, is induced to a delay chain of CARRY4 cells, and the states of CARRY4 outputs are sampled by *Clk160_c* as *RX UsrClk* propagates along the chain. CARRY4 is typically used for fast carry input logic operation in the FPGA as an adder or counter, and is also widely used for time to digital converters [14-17]. A CARRY4 has four cascaded delay cells as shown in Fig. 4. Since the average cell delay in Xilinx Artix-7 FPGA is about 18 ps [14], the total delay of one CARRY4 is about 70 ps. We put 11 consecutive CARRY4 cells in the delay chain and pull two out of four taps in each CARRY4 as the outputs. The whole chain thus covers a time range of about 800 ps, with a time step of about 36 ps. The tap delay in a CARRY4 is not even, and from the work in [17], we pick the first and last as the outputs for better linearity. There are registers following each tap output from the CARRY4 chain to sample the states of the taps (*D[0]-D[21]*). These registers are driven by *Clk160_c*.

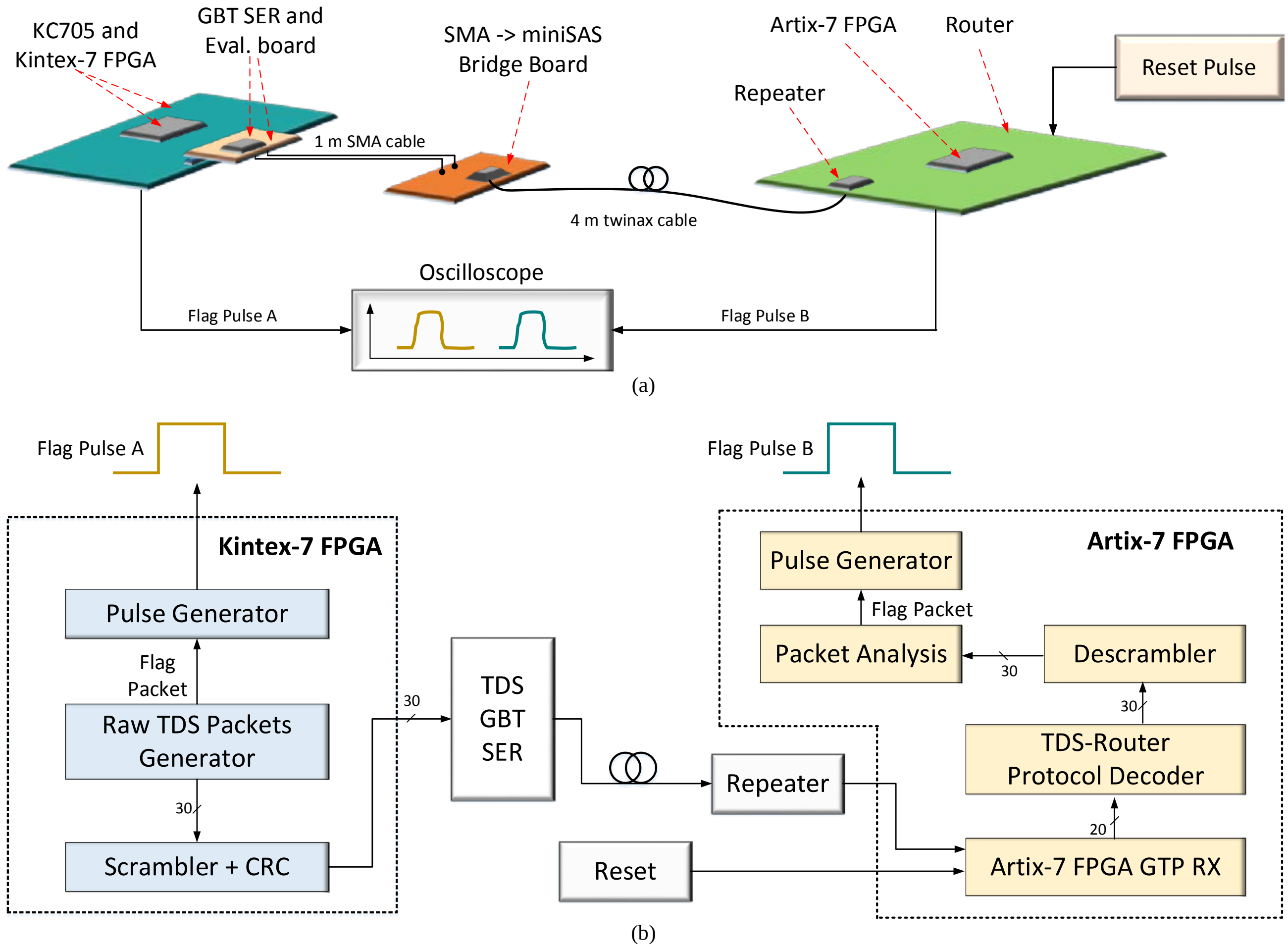

Fig. 6. TDS-Router fixed latency link test setup, in which (a) is a block illustration of the setup; (b) shows the signal flow of the link.

For the rising edge alignment in Step 1, we gradually increase the phase of *Clk160_c* until its rising edge meets that of *RX UsrClk*, which is reflected on the states of CARRY4 chain. The phase of *Clk160_c* can be tuned by selecting the number of delay taps used from the cascaded IDELAYE2 units, and the states of CARRY4 chain are sampled by *Clk160_c* every 16 clock cycles. The phase of *Clk160_c* is forwarded at one step (78 ps) per time and corresponding states of CARRY4 chain are loaded into the “Edge Detector and Control” block for alignment check before another phase step is added. As more taps are enabled in IDELAYE2 units, the rising edge of *Clk160_c* eventually meets the closest rising edge of *RX UsrClk*. This is reflected on the CARRY4 cell status *D[0]-D[21]* switching from an all-zero status “0….0” to a non-zero status “1….0”, as shown in Fig. 5. Tap 0-21 in Fig. 5 corresponds to the 22 delay taps from CARRY4 #0-10 in Fig. 4. As *RX UsrClk* propagates along the CARRY4 chain, *Clk160_c* samples the taps and *D[0]* is the latest state while *D[21]* represents the earliest state of *RX UsrClk*. The samples *D[0]-D[21]* are reversed to correspond to the waveform of *RX UsrClk* in the time domain as shown in both Fig. 5 (a) and (b). Since the range of CARRY4 chain (~800 ps) is less than half cycle of *RX UsrClk*, *D[0]-D[21]* stay in zeroes for some time before the edges are aligned. Once the edges are aligned, *D[0]-D[21]* switch to a non-zero status as shown in Fig.5 (b).

The “Edge Detector and Control” unit monitors *D[0]-D[21]* and once the alignment condition is met, an *Edge Aligned* flag is sent to the “IDELAYE2 CONTROL LOGIC” unit to stop increasing the phase of *Clk160_c* and to record the number of taps used. Meanwhile a *hold* signal is also sent together with *Edge Aligned*, which acknowledges that “Edge Detector and Control” is busy configuring the generation of the *Sample* signal from the aligned clock edges. The rising edges of *RX UsrClk* and those of the aligned 160 MHz clock meet every three cycles of the *RX UsrClk* and the control logic makes sure the *Sample* signal always occur from the rising edge of *RX UsrClk* at each edge alignment point. “IDELAYE2 CONTROL LOGIC” holds its operation until the release of *hold* and then passes the used number of delay taps of IDELAYE2 (*ctrl_r0[9:0]*) to the “IDELAYE2 ctrl register Calculator” for Step 2.

In Step 2, packet header position is derived from the “Syn. & Packet Builder” block as shown in Fig. 3, and “IDELAYE2 ctrl register Calculator” in Fig. 4 converts the frame header position (*head_pos*) into a number of delay taps (*tap_header*). The conversion is done by multiplying *head_pos* by one Unit Interval (UI) of 4.8 Gbps to get the relative time and then divided by 78 ps to get the *tap_header*, i.e. *tap_header* = [*head_pos* × *UI* /*78 ps*], where the operator [*m*] gives an integer closest to *m*. *tap_header* is the amount of phase to be further added to *Clk160_c* from the phase obtained in Step 1

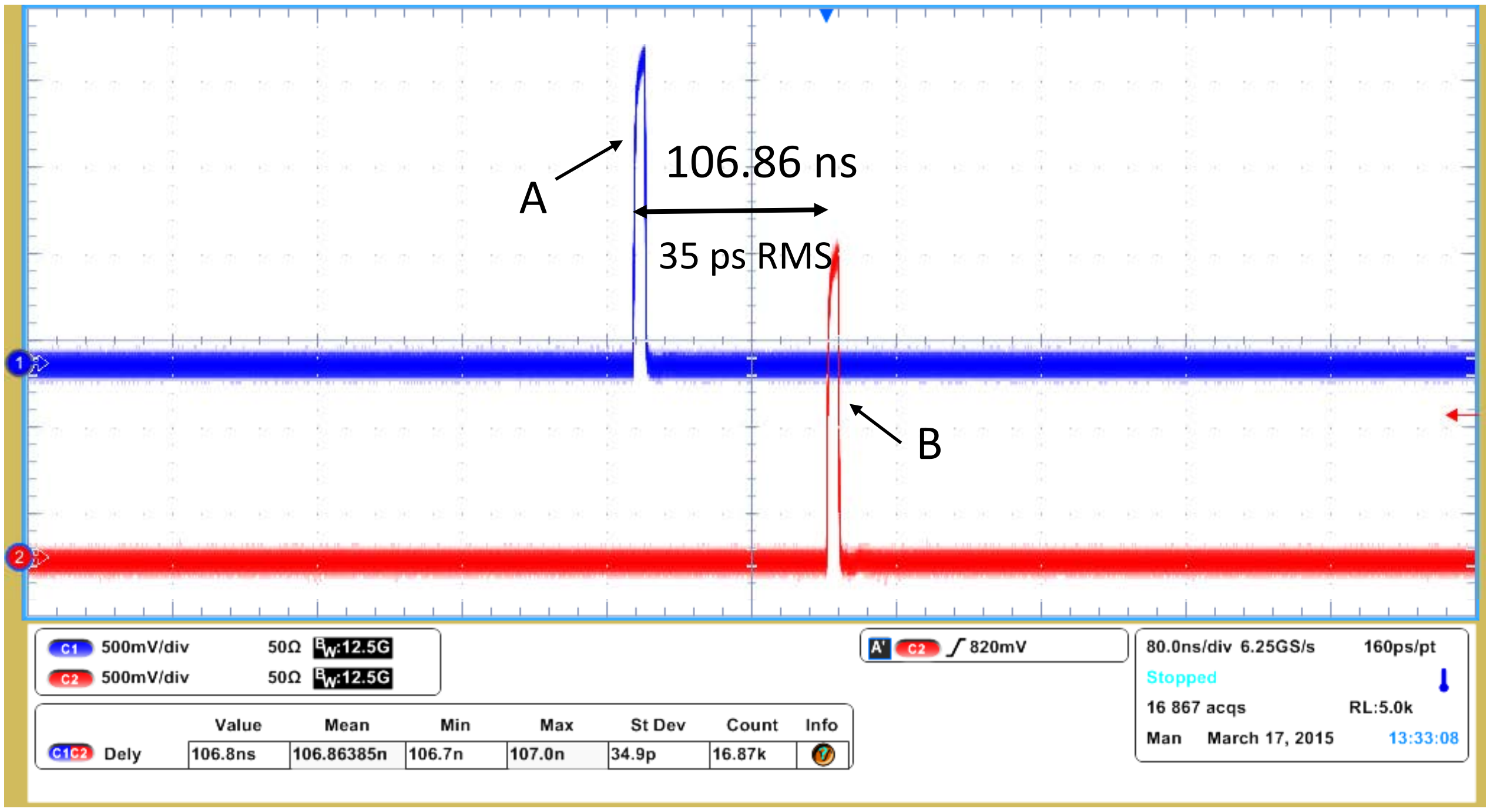

Fig. 7. Typical TDS-Router link latency at a header position of 8 for Router board #0. The diagram is obtained with waveforms persistence opened for pulse A and B in oscilloscope.

(*ctrl_r0 [9:0]*), in together, we get the final configuration taps for IDELAYE2 (*ctrl_r*[14:0]). Getting *Clk160_c* aligned to *RX UsrClk* requires two IDELAYE2 cells (corresponding to a range covering one cycle of *RX UsrClk*, 4.17 ns). Thus there are ten bits of control registers *ctrl_r0[9:0]* with the higher 5 bits for IDELAYE2 #1 and lower 5 bits for IDELAYE2 #0. Similarly, in *ctrl_r[14:0]*, from the MSB to LSB there are the three 5-bit configuration registers for IDELAYE2 #2-0 respectively. It is worth mentioning that if the time from the summation of *tap_header* and *ctrl_r0* is larger than one cycle of *Clk160_c* (6.25 ns), the total taps will round off by 6.25 ns /78 ps ~ 80, as the final configuration parameters to *ctrl_r*.

## III. Results and Analysis

### A. Test Setup

A TDS-Router link has been assembled to evaluate the fixed latency scheme. A block illustration of the setup is shown in Fig.6 (a). On the TDS side, a Xilinx Kintex-7 FPGA evaluation board is used to simulate the TDS logic and a GBT serializer evaluation board [5] is utilized to serialize the TDS logic outputs. The serial output is passed to the Router through one meter of coaxial cable with an SMA connector and 4 meters of twinax cable with a miniSAS connector. The conversion between SMA and miniSAS is made by a “SMA-to-miniSAS” bridge board. On the Router, the serial stream is compensated for transmission loss by a signal repeater (TI *DS100BR410*) before being induced to the GTP transceiver in Xilinx Artix-7 FPGA.

We measure the latency of the link by tracking the time taken by a flag packet travelling from the TDS end to its Router counterpart. A block diagram of the signal flow between the TDS and Router is shown in Fig.6 (b). The FPGA on the KC705 board generates test packets according to the TDS-Router protocol. Every time a flag packet is generated, a pulse (Flag Pulse A) is output to an oscilloscope, as pulse A shown in Fig.7. The raw packets and corresponding cyclic redundancy check (CRC) bits are scrambled to keep DC balanced before being introduced to the serializer (TDS GBT SER) and transmitted at 4.8 Gbps. From the serializer, a serial stream traverses the cables, the repeater, and finally arrives at an Artix-7 FPGA on the Router. Inside the Artix-7 FPGA, the GTP RX transceiver de-serializes the serial stream and prepares the raw bits by 20 bits in parallel at 240 MHz to “TDS-Router Protocol Decoder”. “TDS-Router Protocol Decoder” decodes the serial stream, applies the fixed latency scheme, and organizes the packets by 30 bits per 160 MHz. The content of the packets is recovered by the descrambler, and once a flag packet is recognized a pulse is output to an oscilloscope (Flag Pulse B), as pulse B shown in Fig. 7. In the test, we power up and down the system and also induce a reset pulse from an arbitrary waveform generator to reset the Router. In this way we want to simulate the actual power and reset cycles, and observe the stability of the link latency across these cycles.

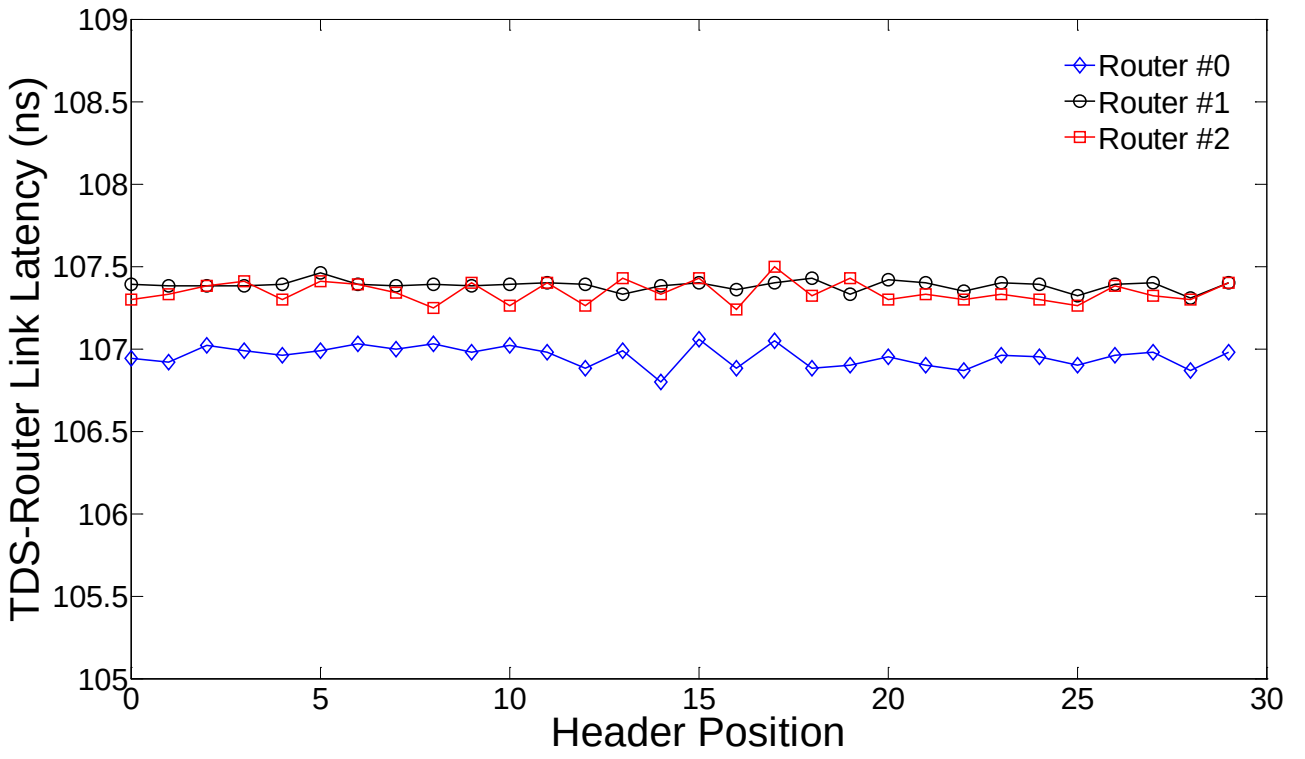

Fig. 8. TDS-Router link latency at all possible header positions for Router board #0-2.

### *B. Results*

The link latency is calculated as the pulse delay between the rising edges of pulse A and B. We perform the tests as follows: we first set the reset period to two minutes, so that at each reset cycle we can accumulate at least 2,000 trials of measurements at a header position. This is repeated until all 30 header positions are observed for a board. Then we carried on the tests for three separate router boards (Router #0, 1, and 2) to test the effect of FPGA process variations on the latency observed. In the test, header positions are monitored from the Xilinx Virtual Input/Output via a JTAG cable. A typical result for the latency is shown in Fig. 7. Embedded statistical analysis in the oscilloscope shows that the average delay is 106.86 ns, with a minimum and maximum value of 106.7 ns and 107 ns respectively, and a variation of 35 ps RMS.

A comparison of the latency for all 30 header positions for Router #0-2 is shown in Fig. 8. The overall latency is found to be flat for all possible header positions in a Router board, with a variation of less than 300 ps. The average latency for all header positions in a Router board also deviates slightly between the three Router boards. Theoretically, the link latency should be identical. The variation most likely comes from the nonlinearity of carry-in taps we used [15-17]. These delay taps are also subject to process variations in the FPGAs. We utilize these delay taps for identification of the clock phase alignment, and the clock phase compensation step is 78 ps, therefore the nonlinearity in tap delay is likely to result in the actual compensated phase value departing from ideal by one or two steps across different packet header positions and FPGAs.

### *C. Breakdown of the Latency*

The latency of the TDS-Router link ($t_{delay}$) needs to be minimal in addition to be deterministic. We analyze it by breaking down the latency into the latency on TDS side ($t_{TDS}$), the latency due to cable delay ($t_{cable}$) and that on the Router side ($t_{Router}$):

$$t_{delay} = t_{TDS} + t_{cable} + t_{Router} \quad (1)$$

In evaluation of the latency, we use identical length of cable for pulse A and B to oscilloscope to avoid any effect on the latency measurement. Contributions to the latency from other parts in the link, such as the propagation delay of the Repeater (~240 ps) is negligible. In (1), the term $t_{TDS}$ is calculated from the generation of raw TDS packets to the first bit out of the GBT serializer. From Fig.6, "Raw TDS Packets Generator" and "Scrambler + CRC" each takes one cycle of 160 MHz clock, therefore, $t_{TDS}$ is the summation of two cycles of 160 MHz clock plus the delay of the TDS GBT SER (8.4 ns, [5]), i.e. the total delay on the TDS side is $t_{TDS}$ = 12.5 ns + 8.4 ns = 20.9 ns. The total length of cable including one-meter SMA coaxial cable and 4 meters twinax is 5 meters. Therefore $t_{cable}$ is roughly 25 ns under the approximation that each one-meter introduces 5 ns delay. As a result, $t_{Router}$ is roughly:

$$t_{Router} = 107 \text{ ns} - 20.9 \text{ ns} - 25 \text{ ns} \approx 61 \text{ ns} \quad (2)$$

In (2), the link latency is taken as 107 ns as shown in Fig.7.

$t_{TDS}$ and $t_{cable}$ are relatively constant, whereas $t_{Router}$ is dependent on the configuration of the GTP transceiver. In our analysis, we further break down $t_{Router}$ into the delay in GTP RX ($t_{RX}$), the time of buffering packets from GTP RX for protocol decoder ($t_{buffer}$) and descrambler ($t_{desc}$):

$$t_{Router} = t_{RX} + t_{buffer} + t_{desc} \quad (3)$$

With our configuration of GTP RX as in Fig. 2, the lowest latency in RX equals to the time of three GTP RX user clock cycles ($t_{UsrClk}$, 240 MHz) plus 123 UI (1 UI ≈ 208.3 ps for 4.8 Gbps) [10], i.e., $t_{RX}$ = 3 × $t_{UsrClk}$+ 123 × 208.3 ps ≈ 38 ns. Protocol Decoder buffers two TDS packets from the 20 bits received from GTP RX, therefore $t_{buffer}$ is one *RX UsrClk* cycle plus 60 UIs, i.e. $t_{buffer}$ = $t_{UsrClk}$+ 60 × 208.3 ps ≈ 16.7 ns. Descrambler takes one cycle of 160 MHz, so $t_{desc}$ = 6.25 ns. If we substitute these analyses into (3) we have $t_{Router}$ ≈ 61 ns, consistent with that in (2).

## IV. Discussion

### *A. Pros and Cons of the scheme proposed*

There are two schemes proposed for fix-latency implementation with FPGA transceivers in [6] [8]. In [6], the internal RX byte and word alignment feature of the FPGA transceiver is utilized, thus the implementation is dependent on this particular feature. Moreover, from [6], the link latency varies from even and odd header positions. A reset circuit is thus needed to keep the link always start with either even or odd header positions for a fixed latency application. The work in [8] releases the dependence on RX byte and word alignment with the phase compensation feature of DCM. However, this scheme will be restricted by the limited number of dedicated clock managing resources inside an FPGA, and will also degrade the performance of other logic as the number of links increases.

The fixed-latency scheme we proposed for Router makes use of carry-input and IO delay resources inside an FPGA. It is independent of internal features of FPGA transceivers, and does not rely on the dedicated clock managing resources (e.g. DCM) either. There is a drawback on this scheme from the non-linearity of tap delay in the CARRY4 chain. In Fig. 8, we evaluated the scheme in FPGAs of three Router prototypes, and observed the link latency vary slightly at different headers and for different FPGAs. The variation is mostly due to the non-linearity of tap delay in the CARRY4 chain.

### *B. Logic Utilization*

A Xilinx Artix-7 FPGA (XC7A200TFFG1156-2) is proposed to be used for the Router. Although the RX byte and word alignment feature is available inside Artix-7 GTP transceivers, the Router RX end needs to synchronize the packets from the domain of *RX UsrClk* (240 MHz) to that of *Clk160_c* (160 MHz) due to the TDS-Router protocol shown in Fig. 3. The fixed-latency scheme with RX byte and word alignment is thus not a good choice as an additional circuit is required to provide an edge synchronized 160 MHz with respect to the phase compensated *RX UsrClk.* Either a clock managing block as DCM is required to generate a 160 MHz from *RX UsrClk* or a special circuit as shown in Fig.4 is needed to align the rising edge of a 160 MHz to that of *RX UsrClk*.

TABLE I
LOGIC UTILIZATION FOR ROUTER RX LOGIC

| Resource | USED/AVAILABLE | Utilization |
|---|---|---|
| Slice Register | 554/269200 | 0.20% |
| Slice LUT | 755/134600 | 0.56% |
| BUFG | 3/32 | 9.37% |
| BUFR | 1/40 | 2.5% |
| DCM/PLL | 0/10 | 0 |
| GTP_DUAL | 1/16 | 6.25% |
| IO_DELAY | 3/500 | 0.6% |

We summarize the logic utilization for one link with the proposed scheme in this paper in Table 1. More links can be instantiated with regard to the available logic resources as the links are independent. We observe that only a very small portion of slice resources with respect to the total available inside the FPGA are used. A total of three global clock buffers (BUFG) are used in our evaluation, one is for the GTP *RX UsrClk*, and the other two are for *Clk160* and a clock for Dynamic Reconfiguration Port ($Clk_{drp}$) of GTP transceivers. In case there are multiple instantiations of GTP channels, *Clk160* and $Clk_{drp}$ can be shared for all links. As a result the number of BUFGs used will only increase by one for every additional GTP RX used. We used a regional clock buffer (BUFR) for the phase compensated 160 MHz clock: *Clk160_c*, and there are a total of 40 available in the Artix-7 FPGA.

## V. CONCLUSION

We propose a new fixed latency serial link scheme for transceivers in FPGAs. The scheme makes use of the FPGA IO delay and carry-in resources. It is protocol independent and offers flexibility for user specific configuration. It also relaxes requirements on dedicated functionality of the transceivers. The scheme is evaluated in a Xilinx Artix-7 FPGA and can be adapted to comparable families of FGPAs in Xilinx or FPGAs from other vendors with similar resources. Our tests from three Router prototypes demonstrate that fixed latency can be achieved at a variation less than 300 ps for all possible header positions. Though the scheme is evaluated in the ATLAS NSW sTGC Router, it can also be adapted to other serial communication applications requiring deterministic latency.